\documentclass[aps,prb,twocolumn,showpacs,floatfix,superscriptaddress,10pt]{revtex4-1}

  \usepackage{float} 
  \usepackage{color}  

\usepackage{graphicx}
\usepackage{amsmath}
\usepackage{amssymb}
\usepackage{bbm}
\usepackage{indentfirst}
\usepackage{dcolumn}
\usepackage{soul}
\usepackage{subfigure}

\DeclareMathOperator{\Tr}{Tr}
\newcommand{\dd}{\,\mathrm{d}}                        
\newcommand{\ttb}[1]{ \small{\texttt{#1}} }           
\newcommand{\unn}[1]{ \underline{\underline{#1}} }    
\newcommand{\mmatrix}[1]{\boldsymbol{\mathrm{#1}}}    
\newcommand{\vrd}{\vec}                               
\newcommand{\mxrd}[1]{\unn{#1}}                       
\newcommand{\mxtd}[1]{\underline{#1}}                 
\newcommand{\mxsd}[1]{\mmatrix{#1}}                   

\begin{document}

\title{Atomistic simulation of finite temperature magnetism of nanoparticles: application to cobalt clusters on Au(111)}
\author{A. L\'aszl\'offy}
\affiliation{Department of Theoretical Physics, Budapest University of Technology and Economics, Budafoki \'{u}t 8., HU-1111 Budapest, Hungary}
\author{L. Udvardi}
\affiliation{Department of Theoretical Physics, Budapest University of Technology and Economics, Budafoki \'{u}t 8., HU-1111 Budapest, Hungary}
\affiliation{MTA-BME Condensed Matter Research Group, Budapest University of Technology and Economics, Budafoki \'{u}t 8, HU-1111 Budapest, Hungary} 
\author{L. Szunyogh}
\affiliation{Department of Theoretical Physics, Budapest University of Technology and Economics, Budafoki \'{u}t 8., HU-1111 Budapest, Hungary}
\affiliation{MTA-BME Condensed Matter Research Group, Budapest University of Technology and Economics, Budafoki \'{u}t 8, HU-1111 Budapest, Hungary} 

\date{\today}

\begin{abstract}
We developed a technique to determine suitable spin models for small embedded clusters of arbitrary geometry by combining the spin-cluster expansion with the relativistic disordered local moment scheme. We present results for uncovered and covered hexagonal Co clusters on Au(111) surface, and use classical Monte Carlo simulations to study the temperature dependent properties of the systems. To test the new method we compare the calculated spin-model parameters of the uncovered clusters with those of a Co monolayer deposited on Au(111).   In general,  the isotropic and DM interactions are larger between atoms at the perimeter than at the center of the clusters. For Co clusters covered by Au, both the contribution to the magnetic anisotropy and the easy axis direction of the perimeter atoms differ from those of the inner atoms due to reduced symmetry.
We investigate the spin reversals of the covered clusters with perpendicular magnetic anisotropy and based on the variance of the magnetization component parallel to the easy direction we suggest a technique to determine the blocking temperature of superparamagnetic particles. We also determine the N\'eel relaxation time from the Monte Carlo simulations and find that it satisfies the N\'eel--Arrhenius law with an energy barrier close to the magnetic anisotropy energy of the clusters.
\end{abstract}

\maketitle

\section{Introduction}
Recent experimental and theoretical  efforts focus on scaling down the size of spintronics and magnetic logics devices to atomic scales to maintain the technological development. 
The superparamagnetic behavior of small ferromagnetic particles gives the size limit of data storage, because the activation energy (energy barrier), $ E_a$ between two stable states of the particle  is proportional to the volume of the particle. The activation energy enters the N\'eel relaxation (average switching) time\cite{Neel1949, Brown1963},
\begin{equation}
	\tau_N = \tau_0 \exp \left( \frac{ E_a}{k_BT} \right),
	\label{eq:neel}
\end{equation}
where $\tau_0$ stands for a characteristic time, $k_B$ is the Boltzmann constant and $T$ is the temperature.  For a given measurement time, $\tau_m$,  the temperature at which only a simple spin flip occurs on average is called the blocking temperature,
\begin{equation}
	T_B = \frac{E_a}{k_B \ln \left( \frac{\tau_m}{\tau_0} \right) }.
	\label{eq:tb}
\end{equation}

The investigations of Co layers and nanoparticles on the surface of Au is a longstanding research subject\cite{Beauvillain1994, Gambardella2003, Miyamachi2014}, with special attention to atomic chains\cite{Gambardella2002}. While a Co monolayer deposited on Au(111) showed in-plane anisotropy, as covered by an additional Au cap an out-of-plane anisotropy has been detected. In addition, the anomalous magnetic anisotropy has been observed in Au/Co/Au(111)\cite{Beauvillain1994} and explained theoretically\cite{Ujfalussy1996}: one monolayer Au coverage induced strong out-of-plane anisotropy, while by increasing the thickness of the Au film the anisotropy decreased, though remained out-of-plane.  Another important observation for Co clusters deposited on Pt(111) was found by  Rusponi {\em et al.}\cite{Rusponi2003}, namely, that the perimeter atoms made significantly larger contribution to the perpendicular magnetic anisotropy (PMA) of the cluster than 
the inner ones, see also Ref.~\onlinecite{kuch2003}. This idea has been explored to produce  nanoparticles with high PMA composed from different $3d$ and $5d$ transition elements and with different geometries\cite{Ouazi2012}. The appearence of large PMA has been pointed out for bcc Co islands on Au(001) by Miyamachi {\em et al.}\cite{Miyamachi2014}, who found a reorianation from in-plane to out-of-plane magnetization with decreasing size of the Co nanoparticles. 



From theoretical point of view, classical spin models are frequently used to study finite temperature magnetism of magnetic nanostructures\cite{Magnetism2007}. To increase the adequacy of such a modelling, the parameters of the spin Hamiltonians can be calculated from first principles. This allows for sorting out the parameters with respect to different atomic positions, which is of crutial importance as indicated above. Embedded cluster techniques combined with the Korringa--Kohn--Rostoker Green's function formalism proved to be extremely useful to study supported small nanoparticles\cite{Wildberger1995, Stepanyuk1999,  Lazarovits2002}. Calculating the exchange interactions between the magnetic atoms in terms of the torque method\cite{Liechtenstein1987} opened the way for atomistic spin-model simulations of such systems\cite{Minar2006, Sipr2007a, Sipr2007b}. The relativistic extension of the torque method (RTM)\cite{udvardi2003,Ebert2009} made it possible to generate an extended spin Hamiltonian including the 
Dzyaloshinsky--Moriya interaction\cite{Dzyaloshinsky1958,Moriya1960} that can induce non-collinear ground state spin-configurations in ferromagnetic nanoparticles\cite{Mankovsky2009, Rozsa2014}.

In this paper we employ an alternative method to calculate the parameters of an extended Heisenberg spin model for embedded clusters. The method relies on the spin-cluster expansion (SCE) originally introduced by Drautz and F\"{a}hnle\cite{Drautz2004}, then extended to the relativistic case as combined with the Relativistic Disordered Local Moment (RDLM) scheme\cite{Szunyogh2011,Deak2011}. A great advantage of the method is that it provides a systematic (irreducible) set of multispin interactions and, once self-consistent potentials and effective fields are provided, the spin-model parameters can uniquely be obtained without the assumption of any arbitrarily ordered reference states. Moreover, the correct symmetry of the exchange interaction and anisotropy matrices is `a priori' granted as dictated by the symmetry of the corresponding lattice site. This is particularly important in case of nanoparticles where different atomic positions, e.g. center or edge positions, have  different  symmetry.

In the next section we briefly describe the SCE-RDLM  method for calculating the spin model parameters of embedded clusters, and also some details of the Monte Carlo simulations we use to study the temperature dependent equilibrium properties of magnetic nanoparticles. Then we show our results for uncovered and covered planar Co clusters on the surface of Au(111). Special attention is paid to the superparamagnetic behavior of the covered clusters with perpendicular anisotropy.

\section{Theoretical and computational details}

\subsection{Embedded cluster technique}

We use the embedding technique based on the Korringa--Kohn--Rostoker (KKR) multiple scattering theory within the framework of density functional theory (DFT) and the local spin-density approximation (LSDA) to determine the magnetic properties of supported transition metal clusters. The details of the method can be found in Ref.~\onlinecite{Lazarovits2002}, here we give only a brief summary.
Within the KKR method the matrix of the scattering path operator (SPO) describing the scattering effects between two of atomic sites for a given energy $ \varepsilon $ is defined as
\begin{equation}
	\mxsd{\tau}(\varepsilon) = \left( \mxsd{t}^{-1}(\varepsilon) - \mxsd{G}_0(\varepsilon) \right) ^{-1},
	\label{eq:mults}
\end{equation}
where $\mxsd{G}_0(\varepsilon)$ is the real space structure constant containing the geometry information and $\mxsd{t}(\varepsilon) = \left\{ \mxtd{t}_i(\varepsilon) \delta_{ij} \right\} $ with $ \mxtd{t}_i(\varepsilon) $ being the single site $t$-matrices. Simple underlines denote matrices in angular momentum space and the bold letters denote matrices in site and angular momentum space, e.g.~$\mxsd{\tau}(\varepsilon) = \lbrace \tau_{ij}^{QQ'}(\varepsilon) \rbrace $ with $i $, $j $ site and $Q$, $Q'$ angular momentum indices, in a relativistic treatment $Q=(\kappa,\mu)$\cite{Rose1961}. To evaluate the $t$-matrices we used the atomic sphere approximation (ASA) with an angular momentum cutoff of $\ell_\mathrm{max}=2$.
%
%
%
%

For an ensemble of magnetic atoms we select a finite environment in which the scattering events are taken into account. The cluster contains not only the magnetic atoms but also a sufficient amount of the perturbed host atoms.  In practice, we first calculate the SPO of the 2D translational invariant layered host within the framework of the screened KKR (SKKR) method, and calculate the $t$-matrices and the SPO matrices confined to the sites of the cluster, $\mxsd{t}_\mathrm{h}(\varepsilon)$ and $\mxsd{\tau}_\mathrm{h}(\varepsilon)$, respectively. The SPO matrix for the embedded cluster, denoted by the subscript $ \mathrm{cl} $, is then evaluated as \begin{equation}
\mxsd{\tau}_\mathrm{cl}(\varepsilon) = \left( \mxsd{\tau}_\mathrm{h}(\varepsilon)^{-1}- \mxsd{t}_\mathrm{h}(\varepsilon)^{-1} + \mxsd{t}_\mathrm{cl}(\varepsilon)^{-1} \right) ^{-1},
\end{equation}
from which the local physical quantities, such as charge and magnetization densities, spin and orbital moments are calculated for the sites of the cluster. In addition, the parameters of an extended Heisenberg spin model can also be determined as described in the next sections.


\subsection{Spin model}
Relying on the adiabatic decoupling of the electronic and spin degrees of freedom and on the rigid spin approximation\cite{Antropov1996} the thermodynamic potential  of a magnetic system is characterized by a set of unit vectors, $ \lbrace \vrd{e} \rbrace = \lbrace \vrd{e}_1 , \dots , \vrd{e}_1 \rbrace $, corresponding to the orientations of the local magnetic moments. The grand potential $ \Omega \left( \lbrace \vrd{e} \rbrace \right) $ then defines a classical spin Hamiltonian which can be used in numerical simulations. Instead of calculating the grand potential directly, a straightforward idea is to map it onto a generalized Heisenberg model in the form:
\begin{equation}
	\Omega\left( \left\{ \vrd{e} \right\} \right) = \Omega_0 + \sum_{i=1}^N \vrd{e}_i \mxrd{K}_i \vrd{e}_i - \frac{1}{2} \sum_{\substack{
   i,j=1 \\
   i \neq j}}^N \vrd{e}_i \mxrd{J}_{ij} \vrd{e}_j,
		\label{eq:heis}
	\end{equation}
where $\Omega_0$ is a constant, $ \mxrd{K}_i $ are the second-order anisotropy matrices and $\mxrd{J}_{ij}$ are the tensorial exchange interactions\cite{udvardi2003}, which can be decomposed into three parts
\begin{align}
	\mxrd{J}_{ij} = & J_{ij}^I\mxrd{I} + \mxrd{J}_{ij}^S + \mxrd{J}_{ij}^A ,
\end{align}
	where 
\begin {align}
	J_{ij} =&\frac{1}{3} \Tr \left( \mxrd{J}_{ij} \right) 
\end{align}
	is the isotropic exchange interaction, 
\begin {align}
	\mxrd{J}_{ij}^S = & \frac{1}{2} \left( \mxrd{J}_{ij} + \mxrd{J}_{ij}^T \right) -J_{ij} \mxrd{I}
 \end{align}
($T$ denoting the transpose of a matrix)	is the traceless symmetric part of the matrix which is known to contribute to the magnetic anisotropy of the system (two-ion anisotropy), and the
	antisymmetric part of the matrix,
\begin {align}
	 \mxrd{J}_{ij}^A = & \frac{1}{2} \left( \mxrd{J}_{ij}-\mxrd{J}_{ij}^T \right) 
 \end{align}
is related to the Dzyaloshinskii--Moriya (DM) interaction,
\begin {align}
 \vrd{e}_i \mxrd{J}^A_{ij} \vrd{e}_j = \vrd{D}_{ij} \left(\vrd{e}_i \times \vrd{e}_j \right)
  \end{align}
  with the DM vector, $D_{ij}^\alpha = \frac{1}{2} \varepsilon_{\alpha \beta \gamma} J_{ij}^{\beta \gamma}$, $ \varepsilon_{\alpha \beta \gamma}$ being the Levi-Civita symbol.

In order to describe the site-resolved magnetic anisotropies, we added the sum of the symmetric part of the exchange matrices to the on-site anisotropy matrix, 
%
%
%
%
%
%
%
%
%
\begin{equation}
\mxrd{A}_i = \mxrd{K}_i - \frac{1}{2}\sum\limits_{\substack{j=1 \\ j\neq i}}^N\mxrd{J}_{ij}^S,
\label{eq:anisite}
\end{equation}
which is still a symmetric matrix. Clearly, for a uniform orientation of the local moments, $\vrd{e}_i=\vrd{e}$, the
energy of the system can be expressed as
\begin{equation}
	\Omega\left( \vrd{e} \right) = \Omega_0 + \sum _{i=1}^N \vrd{e} \mxrd{A}_i \vrd{e} .
	\label{eq:anitot}
\end{equation}
The normalized eigenvectors of the matrix in (\ref{eq:anisite}),  $\vrd{e}^{\,e}_i$, $\vrd{e}^{\,m}_i$, and $\vrd{e}^{\,h}_i$ correspond in order to the  easy, medium and hard directions,  with the respective energies $k^e_i \leq k^m_i \leq k^h_i$. For illustrating the site-specific easy directions together with the magnetic anisotropy energies we will use the following vector, 
\begin{equation}
\vrd{k}^e_i = \left( k^m_i - k^e_i \right) \vrd{e}^{\,e}_i .
\label{eq:ke}
\end{equation}

\subsection{Spin-cluster expansion}
The spin-cluster expansion\cite{Drautz2004} gives a systematic parametrization of the adiabatic magnetic energy of classical spin systems. Restricting ourselves to one-site terms and to pairwise interactions only and using real spherical harmonics, $Y_L\left( \vrd{e}_i \right)$ with the composite angular momentum index $L=(\ell,m)$, the grand potential can be expanded as
\begin{equation}
\begin{aligned}
	\Omega\left( \left\{ \vrd{e} \right\} \right) & \simeq \Omega_0 + \sum_i \sum_{L\neq \left( 0,0 \right) } J_i^LY_L\left( \vrd{e}_i \right) \\
	& + \frac{1}{2} \sum_{i\neq j} \sum_{L\neq \left( 0,0 \right) } \sum_{L'\neq \left( 0,0 \right) } J_{ij}^{LL'}Y_L\left( \vrd{e}_i \right) Y_{L'}\left( \vrd{e}_j \right),
\end{aligned}
\label{eq:sce}
\end{equation}
with
\begin{equation}
	\Omega_0 = \left< \Omega \right>,
	\label{eq:om0}
\end{equation}
\begin{equation}
	J_i^L = \int \mathrm{d}^2e_i \left< \Omega \right>_{\vrd{e}_i}Y_L\left( \vrd{e}_i\right),
	\label{eq:om1}
\end{equation}
and
\begin{equation}
	J_{ij}^{LL'} = \int \mathrm{d}^2e_i \int \mathrm{d}^2e_j \left< \Omega \right>_{\vrd{e}_i \vrd{e}_j}Y_L\left( \vrd{e}_i\right)Y_{L'}\left( \vrd{e}_j\right),
	\label{eq:om2}
\end{equation}
where 
$\left< \quad \right>$ denotes average over all possible spin-configurations, whereas the spin vectors in the subscript, see Eqs.~\eqref{eq:om1} and \eqref{eq:om2}, indicate restricted averages, i.e., we fix the direction of the noted spin vectors and  average with respect to every other spin.
Note that in Eq.~\eqref{eq:sce} the summations do not include the constant spherical function which have the composite index $(\ell,m) = (0,0)$. 
The parameters of the spin Hamiltonian (\ref{eq:heis}) and the SCE coefficients in Eq.~\eqref{eq:sce} can easily be related to each other\cite{Szunyogh2011}. 



%
%
%
%
%
%
%
%
%
%
%
%
%

\subsection{Relativistic disordered local moment scheme}
To evaluate the restricted averages in Eqs.~\eqref{eq:om1} and \eqref{eq:om2} we employed the disordered local moment (DLM) scheme, which was originally introduced as an extension of the conventional spin-density functional theory (SDFT) to include transverse spin fluctuations in the spirit of the adiabatic approximation\cite{Gyorffy1985}. Its relativistic generalization\cite{Staunton2002} can efficiently be used to calculate the spin-model parameters within SCE\cite{Szunyogh2011}. 

Performing averages over spin-orientations requires the evaluation of the single-site $t$-matrices for any spin-direction $ \vrd{e}_i$
which for the case of spherical symmetric potentials (ASA) can be accounted for by the similarity transformation,
\begin{equation}
\mxtd{t}_i\left( \vrd{e}_i \right) = \mxtd{R}\left( \vrd{e}_i \right) \mxtd{t}_i\left( \vrd{e}_z \right) \mxtd{R}\left( \vrd{e}_i \right)^\dag ,
\end{equation}
where $\mxtd{R}\left( \vrd{e}_i \right) $ is the representation of the SO(3) rotation in the angular momentum space ($\dag$ denoting the adjoint matrix) which transforms $\vrd{e}_z$ into $\vrd{e}_i$. Note that the energy argument is not labeled explicitly.

The DLM picture\cite{Gyorffy1985} relies on the coherent potential approximation (CPA) in which an effective (coherent) medium is introduced such that the scattering of an electron is identical as in the original disordered medium on average. This effective medium is represented by the coherent single-site matrices, $ \mxtd{t}_{c,i} $, and the corresponding coherent SPO matrix, 
\begin{equation}
\mxsd{\tau}_c = \left( \mxsd{t}_c^{-1} - \mxsd{G}_0 \right)^{-1},
\end{equation}
or in case of the embedded cluster method,
\begin{equation}
\mxsd{\tau}_{c,\mathrm{cl}} = \left( \mxsd{\tau}_h^{-1}- \mxsd{t}_\mathrm{h}^{-1} + \mxsd{t}_\mathrm{c,cl}^{-1} \right) ^{-1},
\label{eq:tauc}
\end{equation}
The diagonal blocks of $\mxsd{\tau}_c$ satisfy the (single-site) CPA condition,
\begin{equation}
\mxtd{\tau}_{c,ii} = \int \dd^2 e_i  \langle \mxtd{\tau}_{ii} \rangle_{\vrd{e}_i}.
\end{equation}
Defining the excess scattering matrices\cite{Butler1985}
\begin{equation}
	\mxtd{X}_i\left( \vrd{e}_i \right) = \left\{ \left[ \mxtd{t}_{c,i}^{-1}-\mxtd{t}_{i}^{-1}\left( \vrd{e}_i \right) \right] ^{-1} - \mxtd{\tau}_{c,ii} \right\} ^{-1},
	\label{eq:xmatrix}
\end{equation}
the CPA condition can be reformulated as
\begin{equation}
	\int \mathrm{d}^2 e_i \mxtd{X}_i \left( \vrd{e}_i\right) = 0.
	\label{eq:CPA}
\end{equation}
Eqs. (\ref{eq:tauc}), (\ref{eq:xmatrix}) and (\ref{eq:CPA}) can be solved self-consistently to get the coherent single-site $t$-matrices, $ \mxtd{t}_{i,c}$, for each of the magnetic atoms in the cluster.  

In line with the magnetic force theorem used in case of the torque method\cite{Liechtenstein1987,udvardi2003}, Lloyd's formula\cite{Lloyd1967} is used to express the grand potential of the system in the DLM state\cite{Szunyogh2011},
\begin{equation}
\begin{aligned}
& \Omega \left( \left\{ \vrd{e} \right\} \right) = \Omega_c - \frac{1}{\pi} \sum_i\mathrm{Im} \int^{\varepsilon_F}\mathrm{d}\varepsilon \ln \det \mxtd{D}_i \left( \vrd{e}_i \right)\\
& - \frac{1}{\pi} \sum_{k=1}^\infty \frac{1}{k} \sum_{i_1\neq i_2\neq \cdots \neq i_k\neq i_1 } \mathrm{Im} \int ^{\varepsilon_F}\mathrm{d}\varepsilon \Tr \left[ \mxtd{X}_{i_1} \left(\vrd{e}_{i_1}\right) \mxtd{\tau}_{c,i_1i_2} \right. \\
& \left. \times \, \mxtd{X}_{i_2} \left(\vrd{e}_{i_2}\right) \cdots \mxtd{X}_{i_k} \left(\vrd{e}_{i_k}\right) \mxtd{\tau}_{c,i_ki_1}\right],
\label{eq:nagyksz}
\end{aligned}
\end{equation}
where $\Omega_c$ is the configuration independent contribution and 
\begin{equation}
	\mxtd{D}_i\left( \vrd{e}_i \right) = \left\{ \mxtd{I} + \left[ \mxtd{t}_i^{-1}\left( \vrd{e}_i \right) - \mxtd{t}_{c,i}^{-1} \right] \mxtd{\tau}_{c,ii} \right\} ^{-1}.
\end{equation}
is the so-called impuriry matrix.
Using Eq.~\eqref{eq:nagyksz} the restricted averages of the grand potential in Eq.~\eqref{eq:om1} and \eqref{eq:om2} can be calculated.  The onsite SCE coefficients take the form
\begin{equation}
	J_i^L= -\frac{1}{\pi} \mathrm{Im} \int^{\varepsilon_F} \mathrm{d}\varepsilon \int \mathrm{d}^2e_i Y_L \left( \vrd{e}_i \right) \ln \det \mxtd{D}_i \left(\vrd{e}_i \right),
\end{equation}
while, by neglecting backscattering terms\cite{Butler1985}, the pairwise coefficients read as
\begin{equation}
\begin{aligned}
	J_{ij}^{LL'} = &- \frac{1}{\pi} \mathrm{Im} \int^{\varepsilon_F}\mathrm{d}\varepsilon\int \int \mathrm{d}^2e_i\mathrm{d}^2e_jY_L\left( \vrd{e}_i \right) Y_{L'}\left( \vrd{e}_j \right) \\
	& \times \Tr \ln \left[ \mxtd{I} - \mxtd{X}_i \left( \vrd{e}_i \right) \mxtd{\tau}_{c,ij}\mxtd{X}_j \left( \vrd{e}_j \right) \mxtd{\tau}_{c,ji}\right] .
\end{aligned}
\end{equation}

\subsection{Monte Carlo simulations}

Similar to other studies on magnetic nanoparticles\cite{Hinzke1999,Freire2004}, we investigated the temperature dependence of the magnetization by means of classical Monte Carlo simulations using Metropolis algorithm. Assuming that the local magnetic moments vary only a little over the cluster,  the normalized magnetization can be calculated as
\begin{equation}
	\vrd{M} = \frac{1}{N} \sum_{i=1}^N \vrd{e}_{i},
	\label{eq:normmag}
\end{equation}
where $N$ is the number of spins in the cluster.
In absence of external field, the energy of the system is invariant against the reversal of all the spins, therefore, the average 
magnetization of a finite system becomes zero at any temperature.  
We therefore characterize the magnetic system by the absolute value of the average magnetization,
%
%
%
%
%
\begin{equation}
	\left< \left| \vrd{M} \right| \right> = \frac{1}{T} \sum_{t=1}^T \left| \frac{1}{N} \sum_{i=1}^N \vrd{e}_{i,t} \right|,
	\label{eq:tempmagt}
\end{equation}
and by the absolute value of its components,
\begin{equation}
	\left< \left|M_{\alpha} \right| \right> = \frac{1}{T} \sum_{t=1}^T \left| M_{\alpha}  \right| = \frac{1}{T} \sum_{t=1}^{T} \left| \frac{1}{N} \sum_{i=1}^N e_{i,t}^{\alpha}  \right|,
	\label{eq:tempmaga}
\end{equation}
where $t$ labels the measurements and $T$ is the total number of measurements. 
Between two measurements $s$ Monte Carlo steps (MCS) were performed, where one MCS means $N$ Metropolis attempts and {\it s} was chosen typically in order of 10$^4$.  Before taking the averages, the system was thermalized by completing $t_0 \cdot s$ MCS with $t_0 \approx 50$.

%
%

For systems with easy direction ($z$) we found that the deviance of the absolute magnetization in $z$ direction,
\begin{equation}
\sigma_z^2 =  \left< \left( \Delta \left| M_z \right| \right)^2 \right> = \left< \left| M_z \right|^2 \right> - \left< \left| M_z \right| \right>^2,
\label{eq:dev}
\end{equation}
can be used to trace the blocking temperature, $T_B$. In the low temperature limit the magnetization points into $\pm z$ direction, so $\left| M_z \right|$ is practically unchanged, and the deviance approaches to $0$. At larger temperatures some spin flips occur, and the magnetization spend more time in-plane and the deviance of $\left| M_z \right|$ increases with temperature until it reaches a maximum. We found that the deviance temperature, $T_\sigma$, defined as the inflection point of  $\sigma_z^2(T)$, is propotional to the MAE of the system.  Since this applies also to the blocking temperature, see Eq.~\eqref{eq:tb}, the two temperatures can be associated with each other.

\section{Results}

\subsection{Uncovered Co clusters}

\begin{figure}[htb]
	\begin{center}
	\includegraphics[scale=0.18]{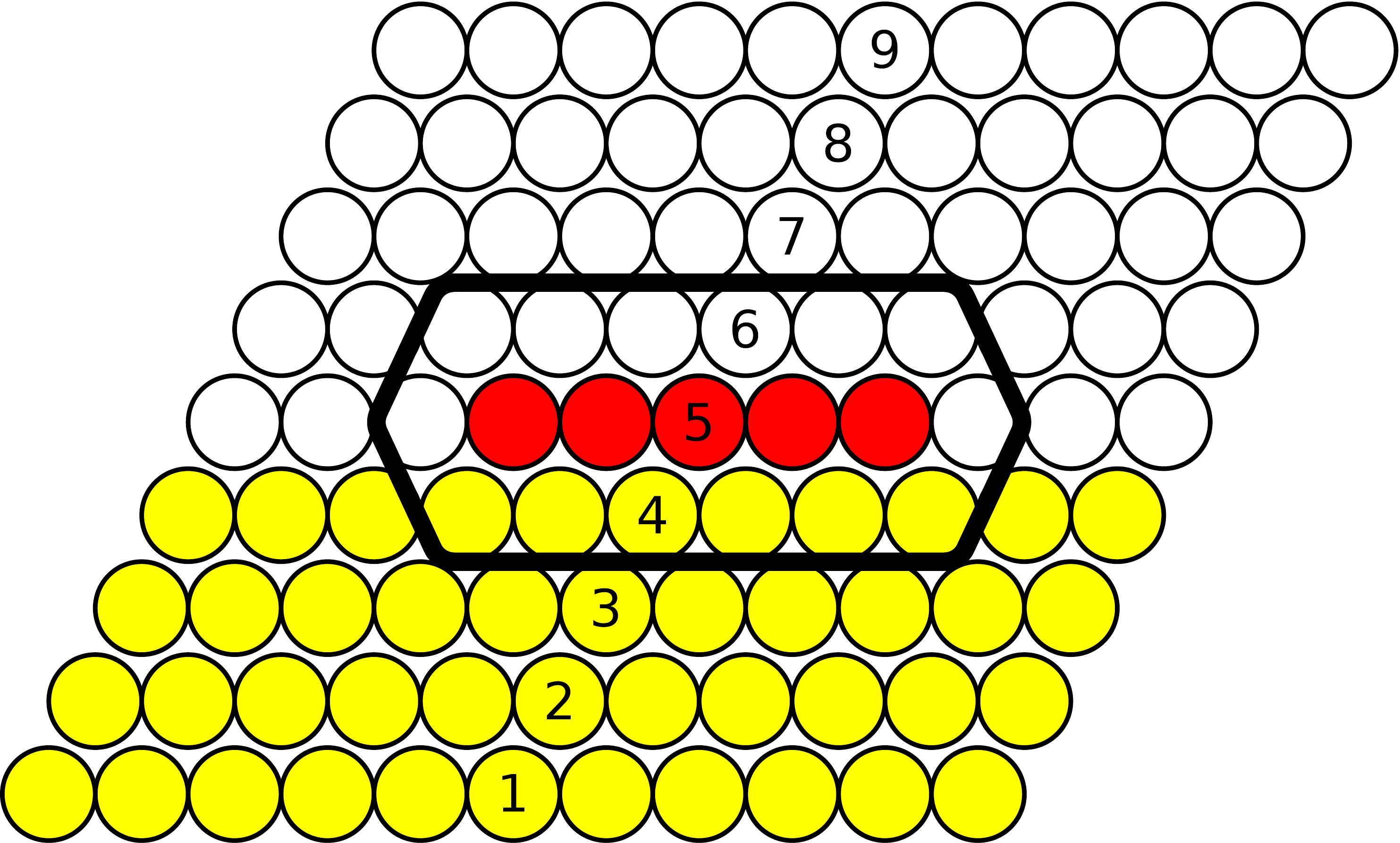}
	\end{center}
	\vspace{-0.5cm}
	\caption{Cross-section illustration of the cluster C2 on Au(111) containing 19 Co atoms. The numbers label the host layers (4 Au layers and 5 empty sphere layers). 
	}
	\label{fig:coxsz}
\end{figure}

We considered three types of planar hexagonal Co clusters deposited on top of the (111) surface of Au, labelled by C1, C2 and C3, and containing 7, 19, and 37 Co atoms, respectively. Each cluster has $C_{3v}$ symmetry, clearly reflected in the calculated magnetic properties. First we perfomed calculations for the (111) surface of Au, where
the topmost four monolayers of Au and five layers of empty spheres (vacuum) were treated self-consistently. 
The cross-section for cluster C2 in Fig.~\ref{fig:coxsz} illustrates how the embedded clusters were contsructed:  
related to both the Au atoms and empty spheres, only those adjacent to the Co atoms were calculated self-consistently. This approach is well justified, since the spin-polarization in Au is quite negligible and, regarding at least the local spin and orbital moments, still reliable in case of Pt substrate with much larger spin-polarization\cite{Lazarovits2003, Sipr2007b}. 

  The self-consistent calculations were performed with ferromagnetic order, with a magnetic orientation perpendicular to the surface ($z$ direction). According to our previous experiences, choosing different global orientations of the magnetization for the self-consistent calculations doesn't remarkably affect the calculated values of magnetic properties.   
 Note that for all systems considered in this work we neglected effects of structural relaxations, i.e.
 both the host  and the embedded atoms occupied positions of a perfect fcc lattice with the lattice constant of bulk Au. This approach allows for investigating pristine effects of the position and the size of the cluster, as well as, the role of the location of atoms within the cluster. 
In order to investigate size effects, we also made calculations for Co monolayer on Au(111).
Note that detailed results will be shown only for cluster C2. 

\begin{figure}[htb]
	\begin{center}
	\includegraphics[scale=0.11]{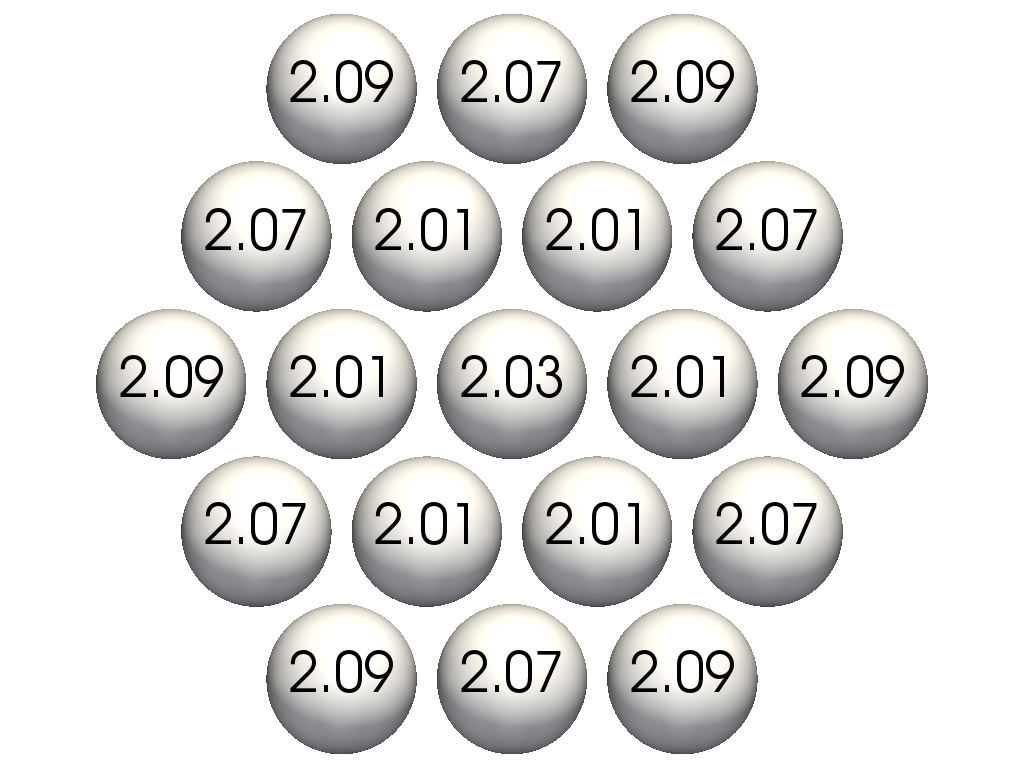} 
	\includegraphics[scale=0.11]{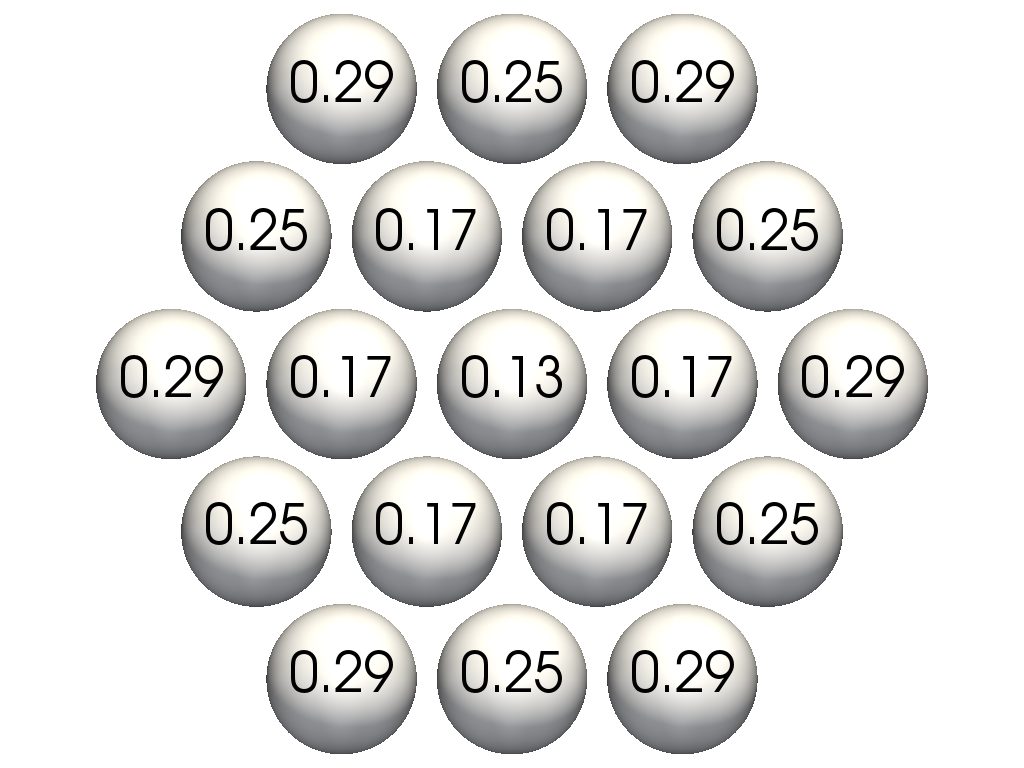}
	\end{center}
	\vspace{-0.5cm}
	\caption{Calculated spin (left) and orbital (right) magnetic moments (in units of $\mu_B$) of the Co atoms in cluster C2.}
	\label{fig:mom}
\end{figure}

The calculated spin and orbital magnetic moments for cluster C2 can be seen in Figure \ref{fig:mom}. The moments for the C1 and C3 clusters are similar to those for C2. As can be seen for the shells with a given distance from the center atom, the magnetic moments connected by a symmetry transformation are the same. The spin-moments are all slightly above  2\,$\mu_B$, and a slight ehancement can be found for the edge and corner atoms (2.07\,$\mu_B$ and 2.09\,$\mu_B$, respectively).   
Owing to different environments of the atoms,  the orbital moments scatter remarkably over the cluster: from 0.13\,$\mu_B$ for the center atom to 0.29\,$\mu_B$ for the corner atoms. Note that our values show great similarity to those reported  for similar clusters in Ref.~\onlinecite{Sipr2007a}. The only remarkable difference is that
in Ref.~\onlinecite{Sipr2007a} the center Co atom in cluster C1 has a spin moment of 1.7\,$\mu_B$, while in our calculations it is 2.02\,$\mu_B$, similar to cluster C2. Considering that the spin and orbital moments of the center atom in cluster C3 are
2.00\,$\mu_B$ and 0.17\,$\mu_B$, the moments approach well the corresponding monolayer values, 1.97\,$\mu_B$ and 0.17\,$\mu_B$, respectively. 

\begin{figure}[htb]
	\begin{center}
	\includegraphics[scale=0.15]{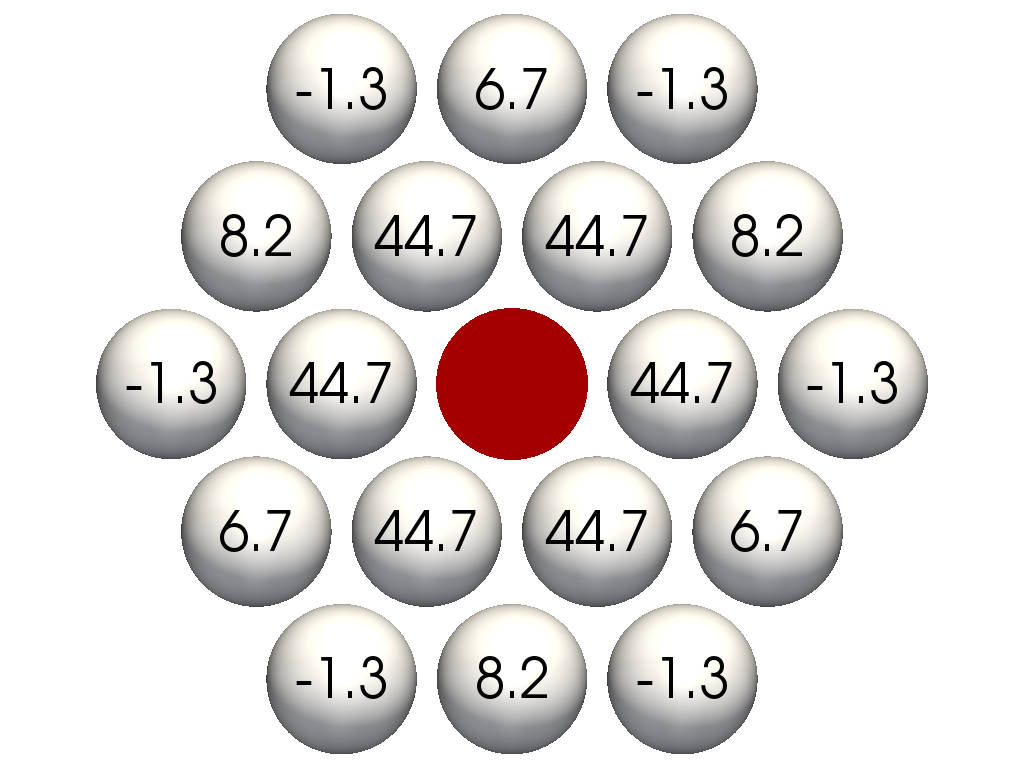}
	\includegraphics[scale=0.15]{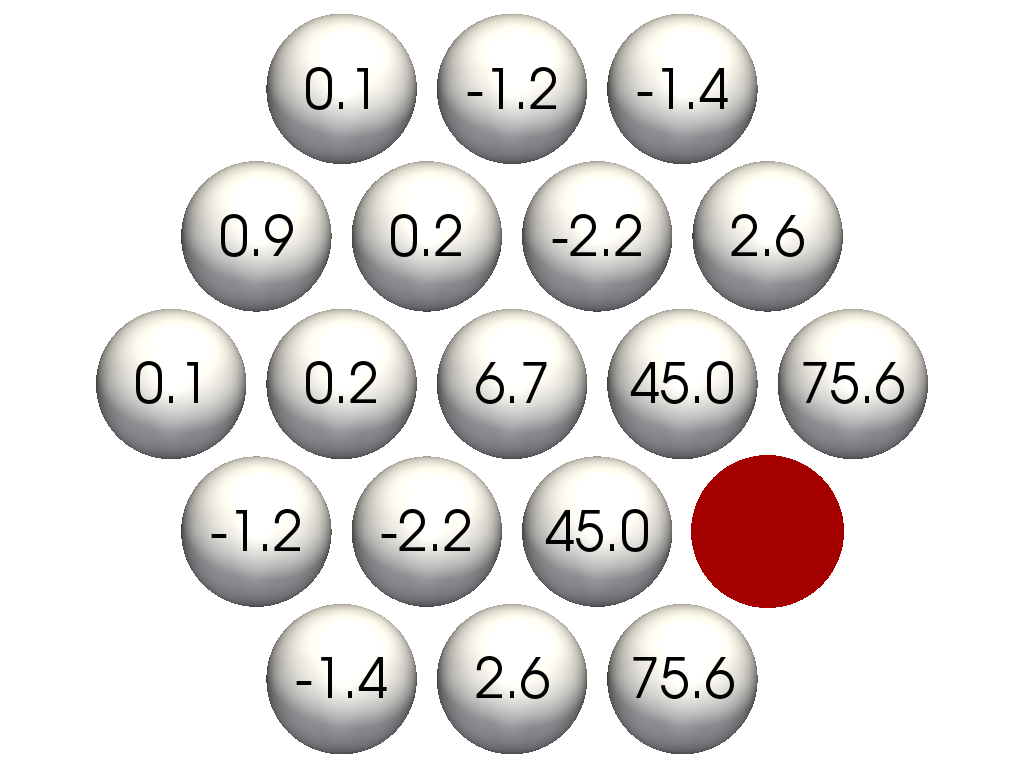}
	\end{center}
	\vspace{-0.1cm}
	\caption{Calculated isotropic exchange interactions (in units of meV) in cluster C2 between the center (left) or edge (right) atom (colored in red) and all the other Co atoms.}
	\label{fig:iso}
\end{figure}

Next we calculated the tensorial exchange interactions and on-site anisotropy matrices by using the SCE-RDLM method described in the previous section. The first nearest neighbor (NN) isotropic interactions are strongly ferromagnetic (positive) and vary between $36.8 - 75.6\,\mathrm{meV}$, while the second neighbor couplings are by about one order smaller, $-7.9 - 8.2\,\mathrm{meV}$. The interaction between the outer atoms are significantly larger because of their reduced coordination, i.e.,  less magnetic neighbor atoms.
The isotropic interactions for the edge and corner atoms of the cluster C2 are shown in Figure \ref{fig:iso}.  The symmetry relationships are clearly recovered in the interactions. Apparently, the interaction between the adjacent edge and corner atoms are largely enhanced due to the reduced coordination of both types of atoms. 

A direct comparison can be made for cluster C1 (1 center atom and 6 perimeter atoms) to the $J_{ij}$ values reported in Ref.~\onlinecite{Bornemann2012}. Though an overall good agreement can be found, the interactions calculated in terms of SCE in this work are  by about 20 \% larger than those obtained from the torque method in  Ref.~\onlinecite{Bornemann2012}.
This can also be seen in the effective exchange field, $J_i = \sum_{j (\ne i)} J_{ij}$, which takes 248\,meV and 199\,meV by the SCE, while    209\,meV and 150\,meV by the torque method\cite{Bornemann2012} for the center and the perimeter atoms, respectively.

\begin{figure}[H]
	\begin{center}
	\includegraphics[scale=0.08]{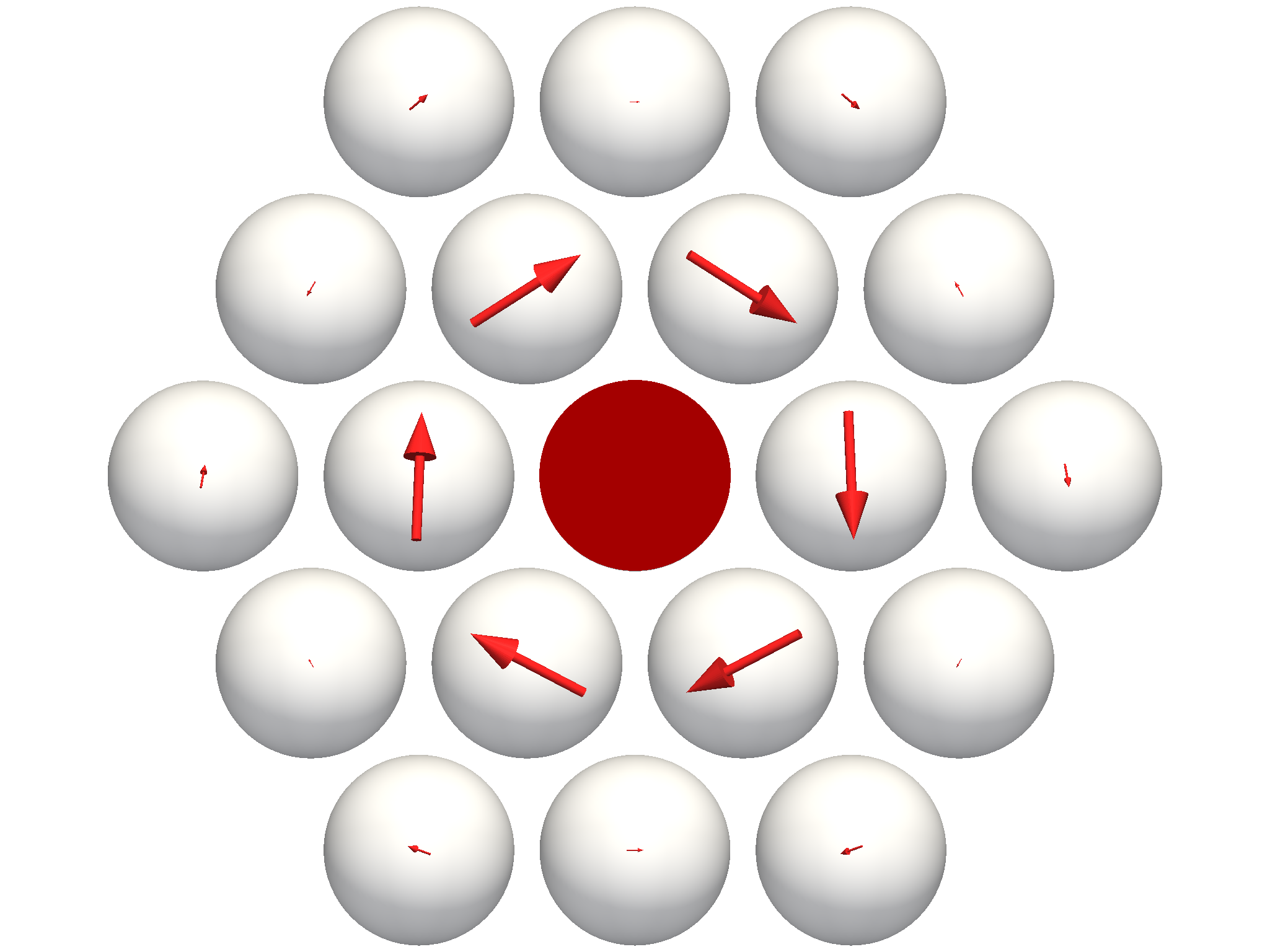}
	\hspace{-0.3cm}
	\includegraphics[scale=0.08]{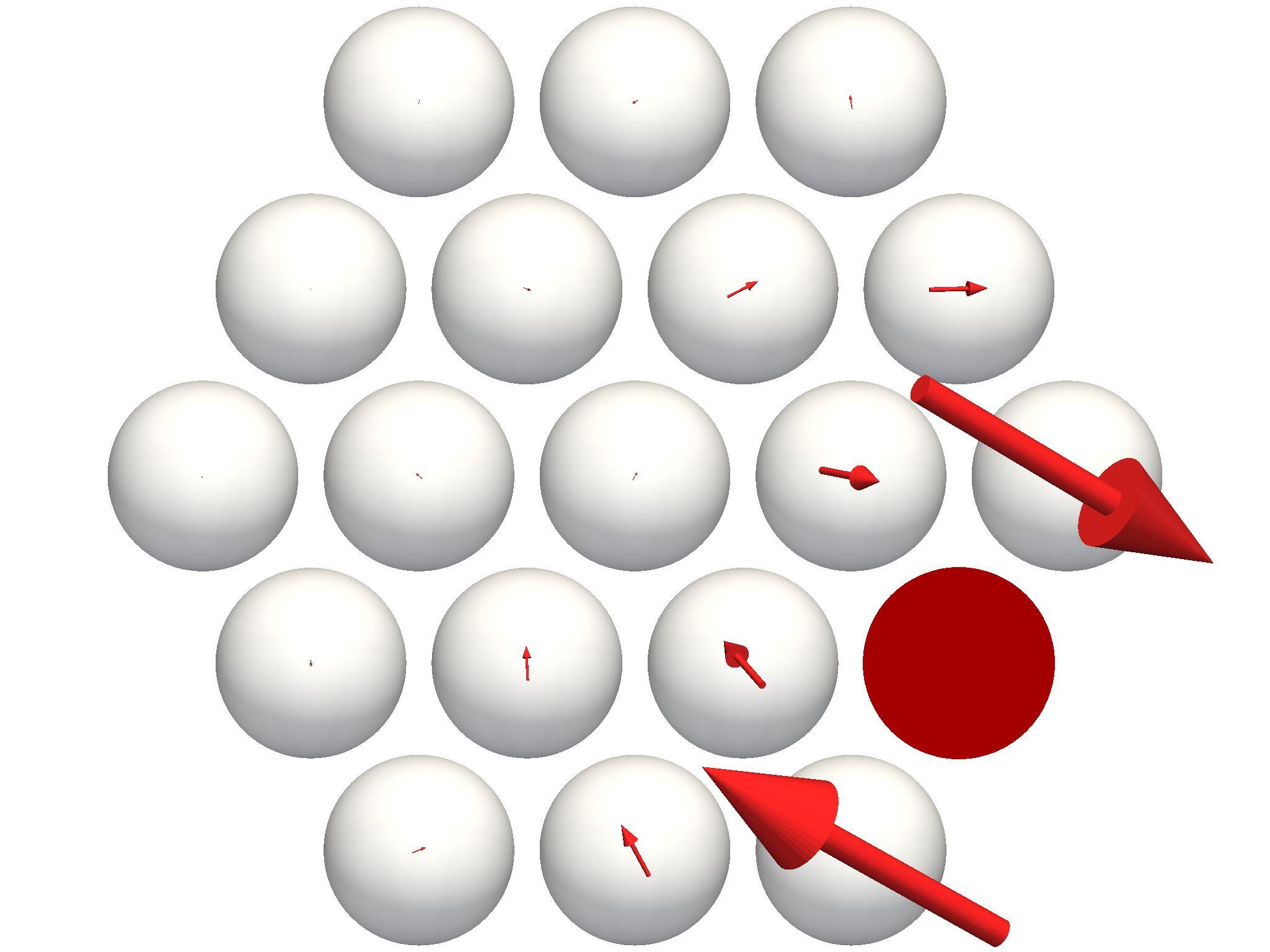}
	\end{center}
	\vspace{-0.1cm}
	\caption{Top view of the calculated DM vectors in cluster C2 between the center (left) or edge (right) atom and all the other Co atoms.}
	\label{fig:dm}
\end{figure}

The magnitudes of the Dzyaloshinkiy--Moriya vectors are typically one order smaller than those of the isotropic interactions, reaching a maximum value of  $2.19$, $4.14$ and $4.47\,\mathrm{meV}$ in the C1, C2 and C3 clusters, respectively. The DM vectors between the center atoms and their nearest neighbor site are around 1\,meV in size, while the NN DM vectors at the rim of the clusters are about 3-4 times larger. This is presented in Fig.~\ref{fig:dm} for the cluster C2.  The orientations of the DM vectors should be assessed taking into account that they behave as axial vectors: in case of reflection symmetry the component parallel to the mirror plane turns round, and the perpendicular component remains unchanged.
Similar to the Co/Au(111) monolayer\cite{Beauvillain1994,Ujfalussy1996}, the considered hexagonal clusters have easy-plane uniaxial magnetic anisotropy in the ferromagnetic state  with and average MAE per Co atom of $0.078\,\mathrm{meV}$ for C1, $0.266\,\mathrm{meV}$ for C2 and $0.596 \,\mathrm{meV}$ for C3, i.e., about two order smaller than the NN isotropic interactions.

	\begin{table}[htb]
	\begin{center}
	\begin{tabular}{|r|rrrr|}
	\cline{2-5}
	\multicolumn{1}{c|}{ } &\multicolumn{1}{c}{C1} &\multicolumn{1}{c}{C2} &\multicolumn{1}{c}{C3} &\multicolumn{1}{c|}{ML}\\
	\hline
	$J_{ij}^I$	& \ttb{41.33}	& \ttb{44.69}	& \ttb{41.28}	& \ttb{36.91}	\\
	$\mathrm{D}_{ij}^x$	& \ttb{-0.043}	& \ttb{0.058}	& \ttb{0.025}	& \ttb{0.000}	\\
	$\mathrm{D}_{ij}^y$	& \ttb{-0.478}	& \ttb{-1.341}	& \ttb{-1.533}	& \ttb{-1.246}	\\
	$\mathrm{D}_{ij}^z$	& \ttb{0.867}	& \ttb{0.389}	& \ttb{0.430}	& \ttb{-0.132}	\\
	$| \vrd{\mathrm{D}}_{ij} |$	& \ttb{0.991}	& \ttb{1.397}	& \ttb{1.592}	& \ttb{1.253}	\\
	\hline
	\end{tabular}
	\caption{Calculated isotropic interactions and DM vectors between the center atom and its first neighbor along the $x$ axis in the three Co clusters and in the Co monolayer on Au(111). All values are given in units of meV.}
	\label{tab:khc13mon}
	\end{center}
	\end{table}
	
In Table \ref{tab:khc13mon} we investigate how the NN interactions at the center of the cluster evolve by increasing the size of the system. As can be seen the interactions do not change dramatically, but the size of the clusters are apparently too small to show a straight convergence to the corresponding monolayer values. A precise convergence is not expected at all, since in case of the monolayer calculation, beside the Co monolayer, four-four monolayers of Au and empty spheres were treated self-consistently, while, as mentioned before, in case of the cluster calculations this applied only to the Au atoms and empty spheres adjacent to the Co atoms. Noticeably, in case of the monolayer the $x$ component of the DM vector vanishes by symmetry for the NN pair along the $x$ axis. Since the center atom and its first neighbor are not connected by any symmetry operation in the clusters, the $x$ component of the DM vector remains finite and it is expected to vanish only in the limit of the monolayer. 

Due to the large ferromagnetic NN isotropic interactions and easy-plane magnetic anisotropy, from the Monte Carlo simulations we obtained a nearly collinear ferromagnetic ground state with the spins pointing parallel to the plane. We observed only a small deviation from collinearity due to the DM interactions. Note that the ground state of these systems is continuously degenerate, since according to the model (\ref{eq:heis}) there is no preferred direction within the plane in case of uniaxial ($C_{3v}$) anisotropy.

\begin{figure}[htb]
	\begin{center}
	\includegraphics[scale=1]{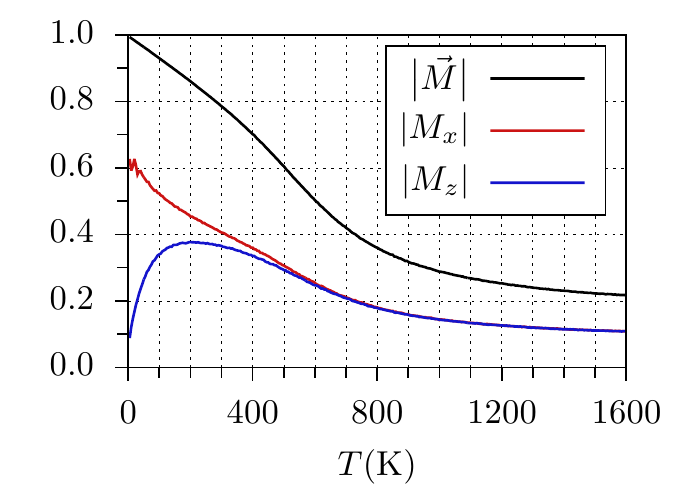}
	\end{center}
	\vspace{-0.5cm}
	\caption{Average magnetization and its components for cluster C3 as defined in Eqs.~\eqref{eq:tempmagt} and \eqref{eq:tempmaga}, respectively.}
	\label{fig:tempmag}
\end{figure}

Because in the considered systems the local magnetic moment varied only very little from site to site, see Fig.~\ref{fig:mom}, we calculated the normalized magnetization by Eq.~\eqref{eq:normmag}, and the temperature dependent average magnetization by Eqs.~\eqref{eq:tempmagt} and \eqref{eq:tempmaga}. In Fig.~\ref{fig:tempmag} we show the temperature dependence of these quantities for cluster C3. For the MC simulations we used the parameters $T=40000$, $t_0=50$, and $s=40000$. We used only the half of the sphere to generate the new direction of the random spin (centered to its original direction) to avoid a large number of abortive simulation attempts\cite{Hinzke1999}. In the low temperature limit the magnetization $| \vrd{M} |$ converges to $1$ (in fact, to a slightly smaller value because the ground state is not perfectly collinear). Because of the easy--plane anisotropy, $|\vec{M} | \rightarrow 0$  and $| M_x | \rightarrow 2/\pi\approx 0.637$ which is obtained by $ \displaystyle \frac{1}{2\pi} \int_0^{2\pi} \left| \sin\left( \phi \right) \right| \mathrm{d}\phi $ due to the continuously degenerate ground state. 
In the high temperature limit $| \vec{M} |$  converges to $1/\sqrt{N}$ following from Eq.~\eqref{eq:tempmagt} using independent, uniform distribution to the spin directions. Its components converge to half of it because calculating the expectation value of a component's absolute value with uniform directional distribution is just the same as getting the centroid of a hemispherical shell. $| \vec{M} |$ decreases monotonously with temperature and its inflection point is related to the strength of isotropic interactions. Due to the in-plane anisotropy of the cluster we find $| M_x | > \left| M_z \right|$ at any temperature, however, beyond a certain temperature, which is related to the anisotropy energy, the two components take practically the same values. 
The clusters C1 and C2 show similar behavior, but the temperature where the in-plane and out-of-plane component of the magnetization become the same is shifted to smaller temperatures because their anisotropy energy is much smaller than that of cluster C3.

\subsection{Capped Co clusters}

As indicated by the monolayer case experimentally\cite{Beauvillain1994} and in theory\cite{Ujfalussy1996},  similar Co clusters but covered by gold are supposed to show strong perpendicular magnetic anisotropy that might be of considerable interest for applications. Therefore, we focused our studies on the superparamagnetic behavior of such nanoclusters.  

	\begin{figure}[htb]
		\begin{center}
		\subfigure[\ L3C1]{
		\includegraphics[scale=0.18]{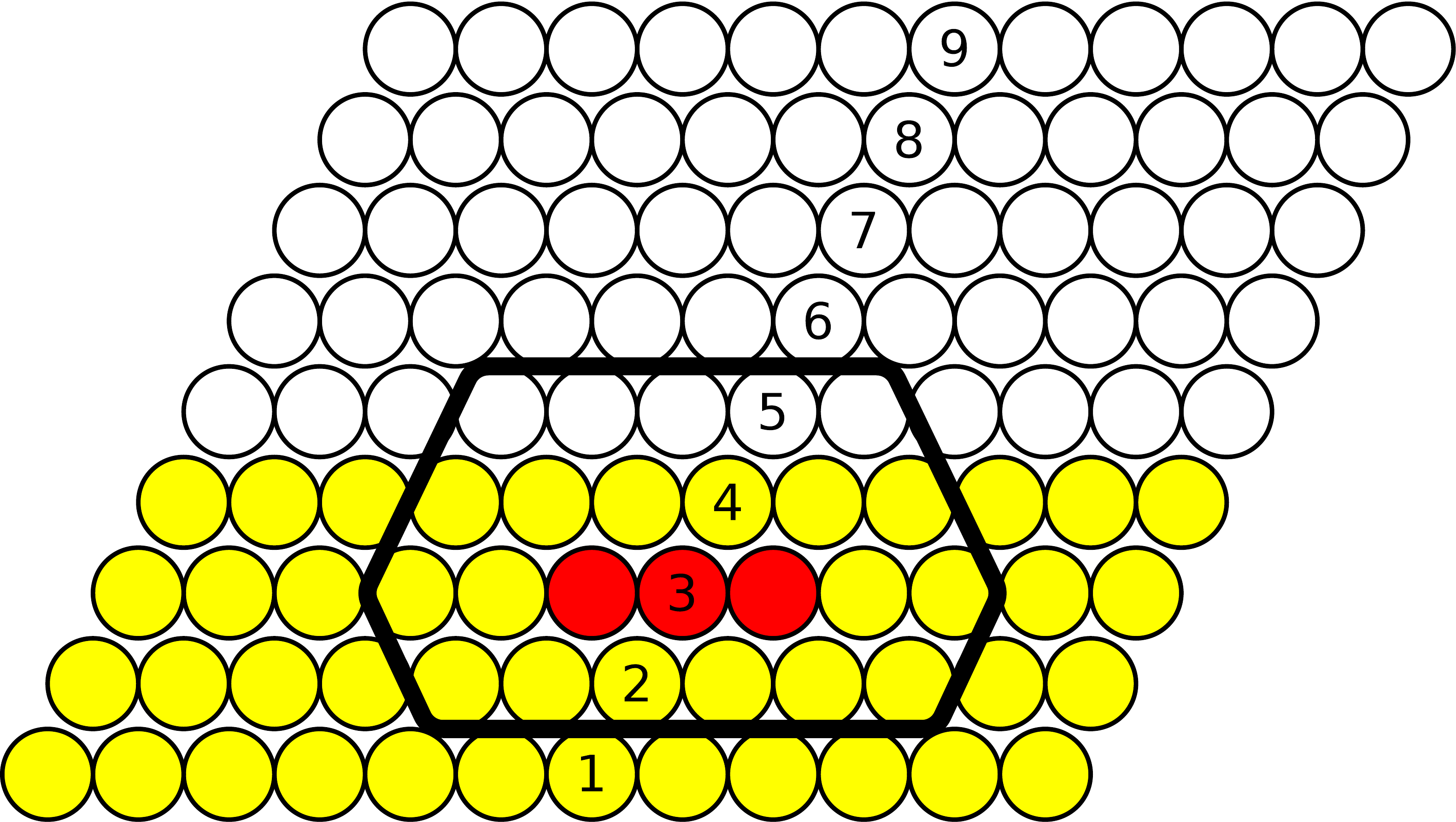}
		}
		\vspace*{-0.15cm}
		\hspace{-1.5cm}
		\subfigure[\ L4C1]{
		\includegraphics[scale=0.18]{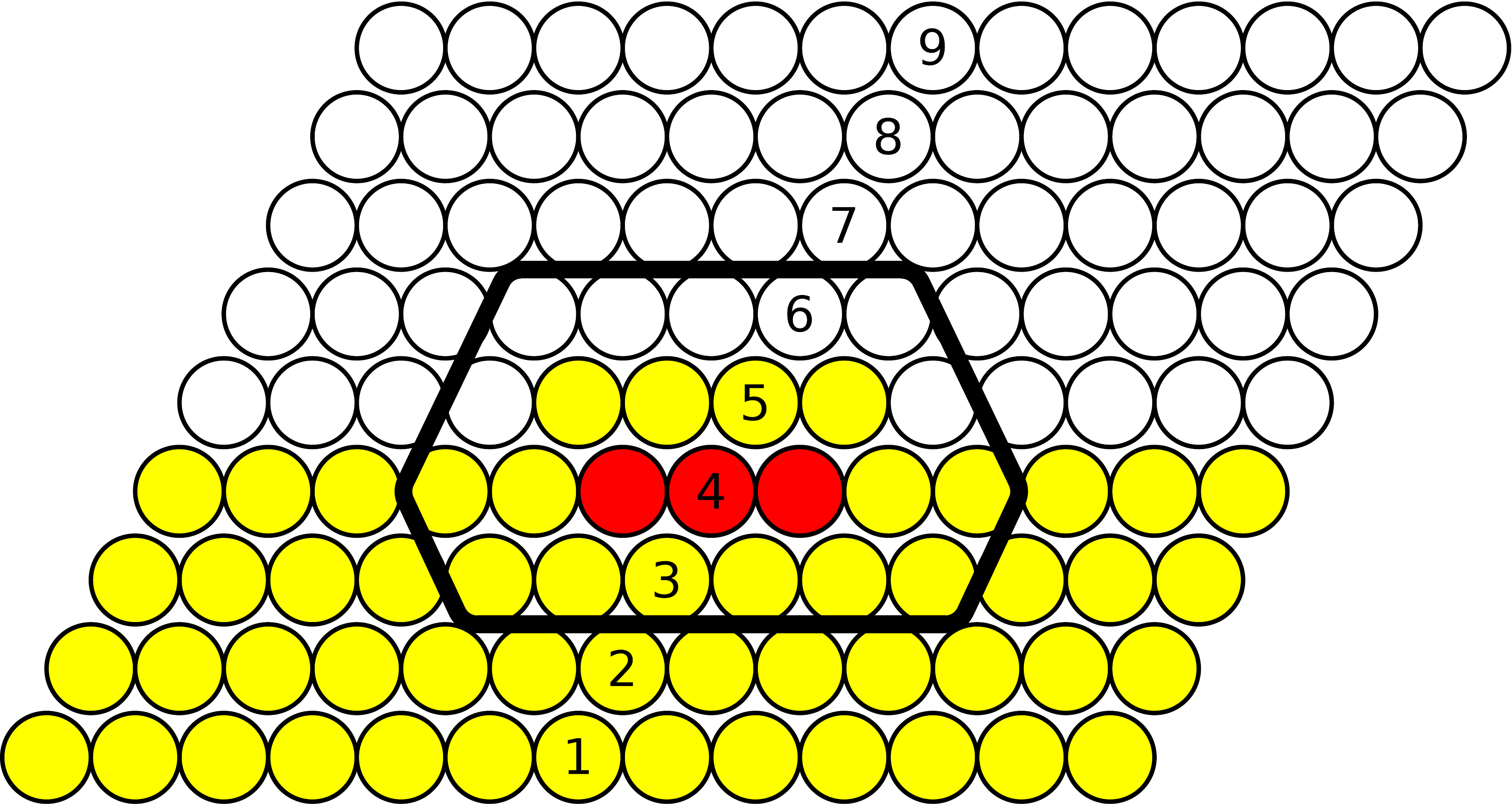}
		}
		\vspace*{-0.15cm}
		\hspace{-1.5cm}
		\subfigure[\ L5C2]{
		\includegraphics[scale=0.18]{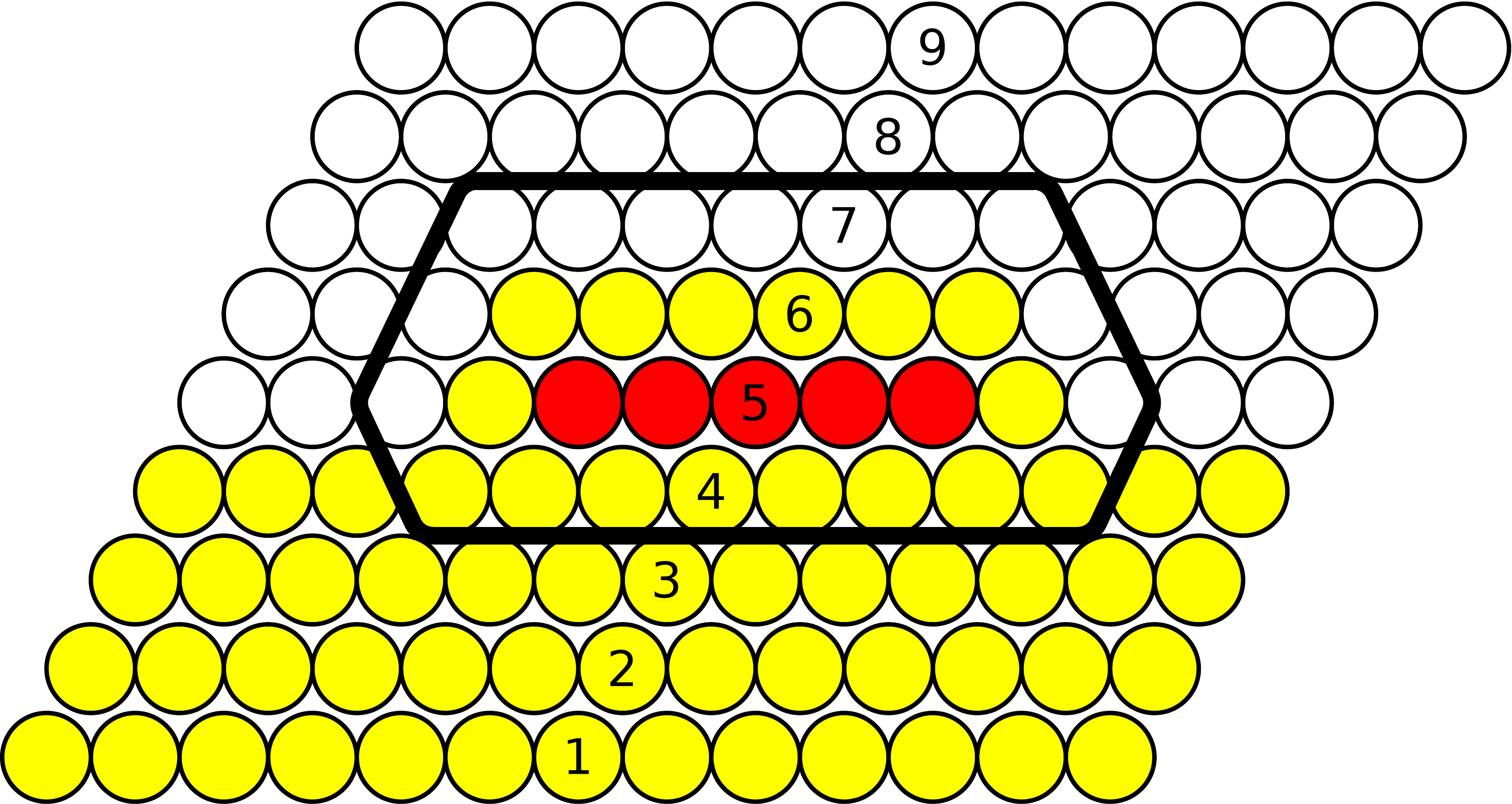}
		\label{fig:cll5c2}
		}
		\end{center}
		\vspace{-0.5cm}
		\caption{Cross-section illustrations of planar Co clusters on Au(111) covered by Au. The number after L denotes the index of the host layer (the numbering of the layers is also presented) the Co atoms are embedded into, and the number after C corresponds to the size of Co cluster (1: 7 atoms, 2: 19 atoms). 
		 }
		\label{fig:laycox}
	\end{figure}

Some of the planar clusters we calculated are shown in Fig.~\ref{fig:laycox} and labeled by LxCy, where $\mathrm{x} \in \{3,4,5\}$ is the label of the host layer the Co atoms are embedded (5: first empty sphere layer, 4: topmost Au layer, 3: subsurface Au layer) and $\mathrm{y} \in \{1,2\} $ corresponds to the size of the cluster similarly to the uncovered case. Contrary to the uncapped Co clusters, the second neighbor empty spheres and sufficiently more Au atoms are included in the clusters. This allows for more precise calculations needed, in particular, for the PMA induced by the gold coverage. 

The NN isotropic interactions are about $20\%$ smaller than for the uncovered clusters and they show little sensitivity to the layer position of the cluster. The magnitudes of the DM interactions are below $2.9 \,\mathrm{meV}$ for all clusters, so they cause only little deviations from a collinear configuration in the ground state. Nevertheless, we found that the DM vectors change drastically, both in direction and in magnitude, when changing the embedding layer.  This can be attributed to the fact that the DM interactions are induced by spin-orbit coupling, therefore, must be strongly influenced by the environment of the cluster formed by the Au atoms.

\begin{figure}[htb]
		\begin{center}
		\vspace*{-0.17cm}
		\subfigure[\ L3C2]{
		\includegraphics[scale=0.05]{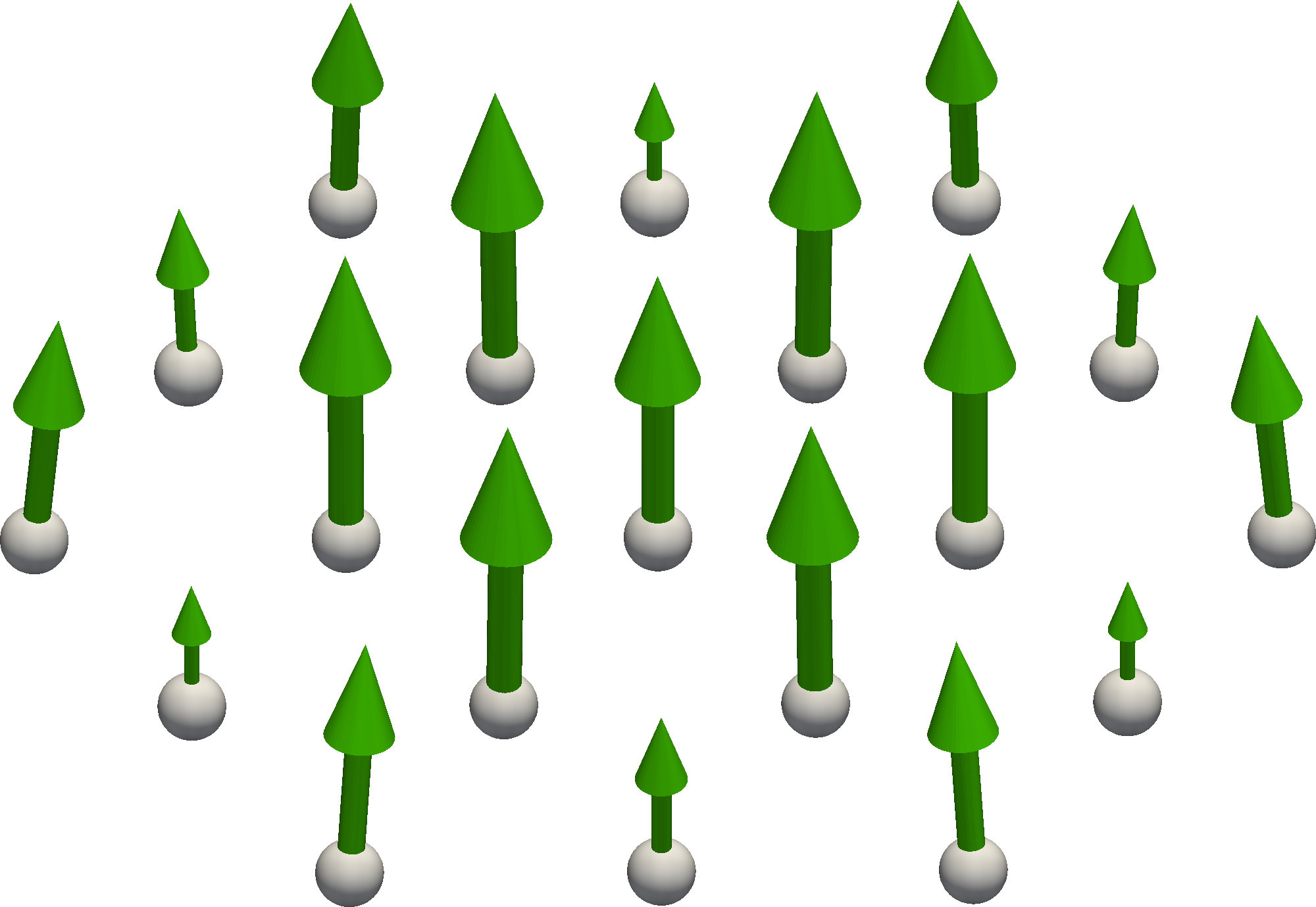}
		}
		\vspace*{-0.13cm}
		\hspace{0.5cm}
		\subfigure[\ L4C2]{
		\includegraphics[scale=0.05]{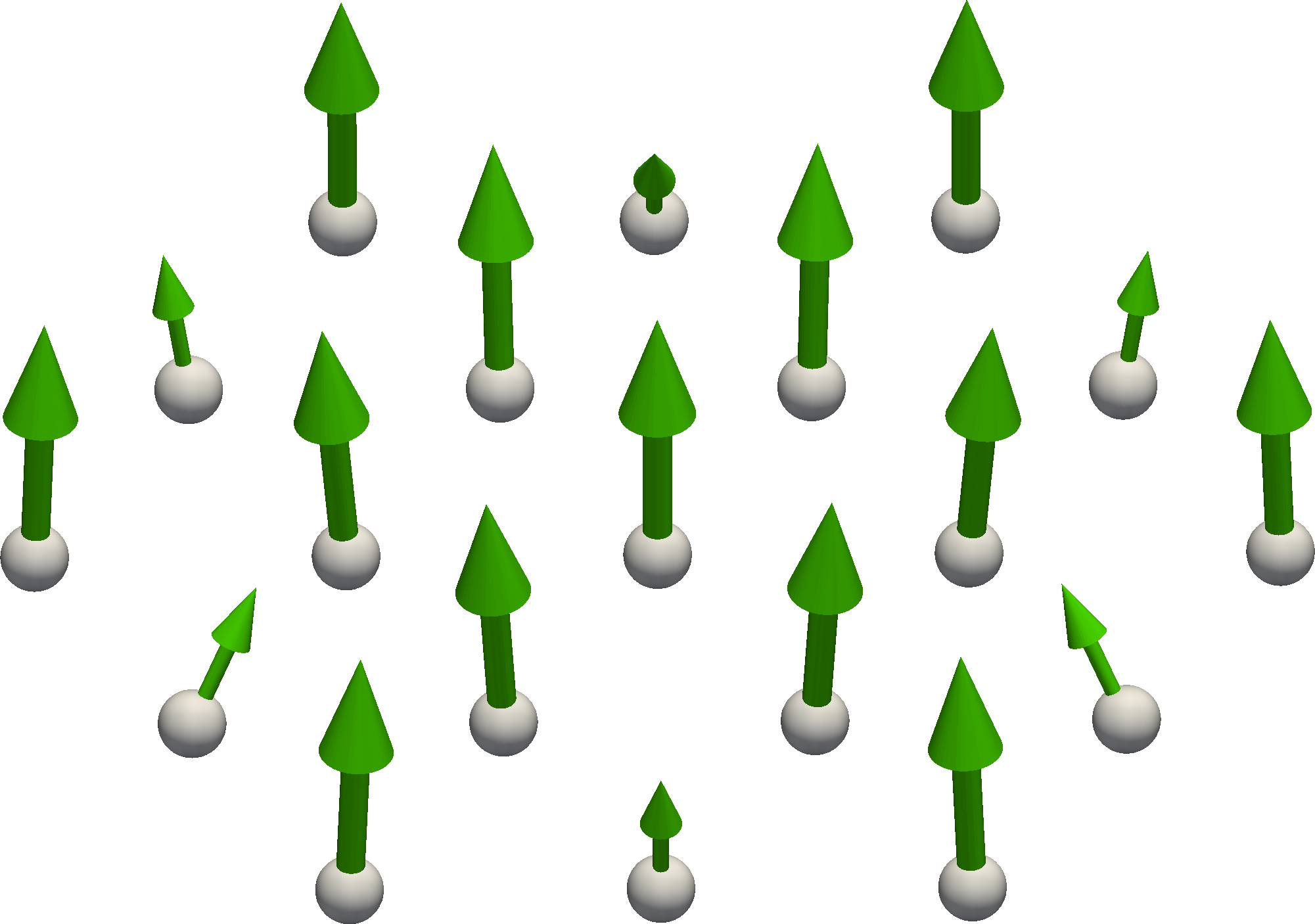}
		}
		\vspace*{-0.13cm}
		\subfigure[\ L5C2]{
		\includegraphics[scale=0.05]{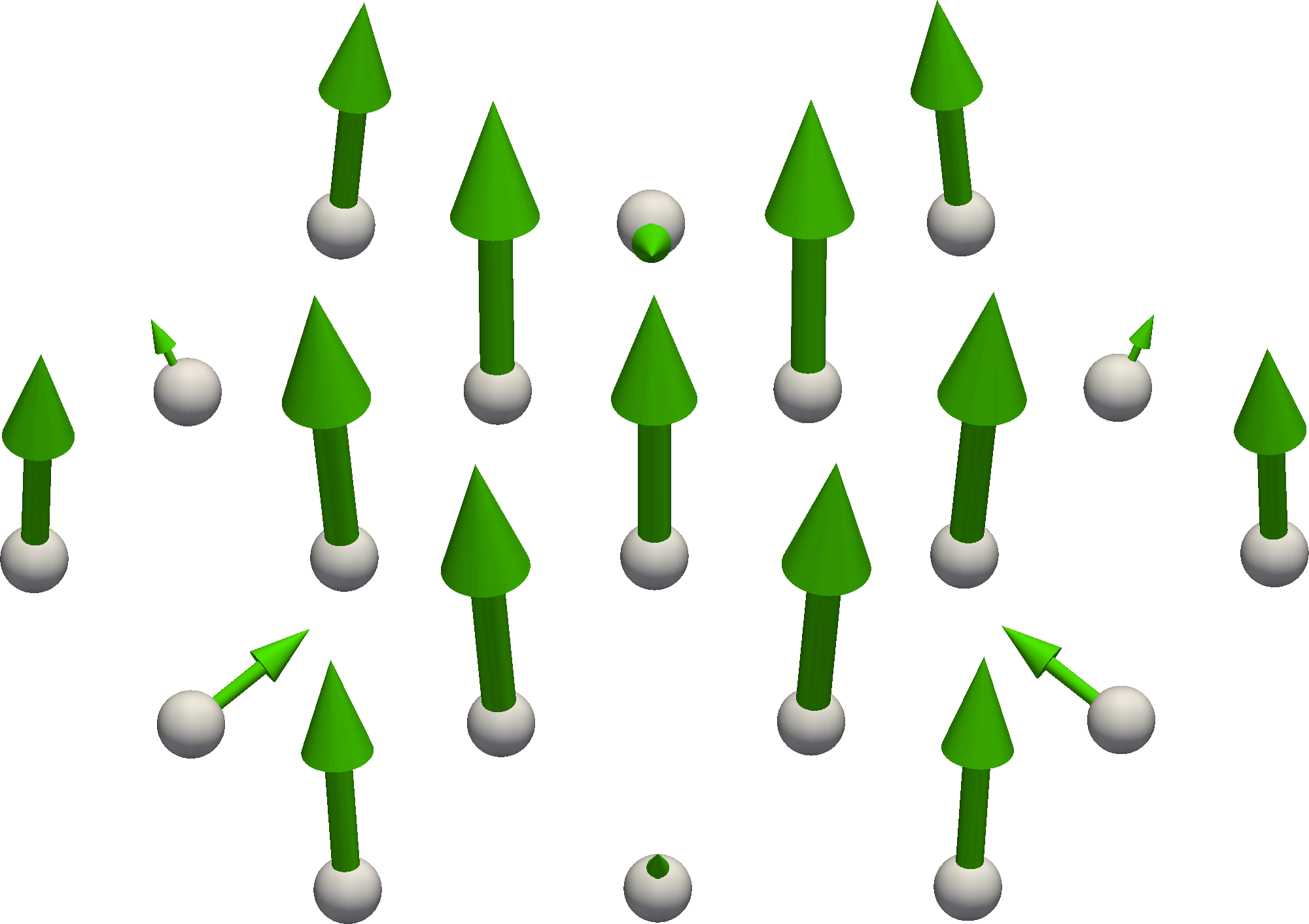}
		}
		\end{center}
	\vspace{-0.5cm}
	\caption{Site-resolved magnetic anisotropy vectors in cluster L4C2 according to Eq.~\eqref{eq:ke}. The vectors are parallel to the easy direction and their length is propotional to the energy difference between the easy and medium directions. 
	}
	\label{fig:ani}
\end{figure}

In Fig.~\ref{fig:ani} the site-resolved magnetic anisotropy vectors as defined in Eq.~\eqref{eq:ke} are presented for clusters L3C2, L4C2 and L5C2. As predicted,  for most of the sites the easy direction is close to being perpendicular to the surface. 
Interestingly,  the largest deviation from uniaxial anisotropy is found for the edge atoms as their easy axis have the largest in-plane component. 
In general, the Co clusters embedded fully into the Au substrate, i.e., the clusters L3Cy show definite out-of-plane anisotropy at each site, but placing the clusters into the surface layer  the easy axes of the corner atoms of the small clusters (L4C1) and of the edge atoms of the large clusters (L4C2)  are tilted with respect to the $z$ diretion. In case of the clusters on top of the surface (L5Cy) the easy axes for these atoms turn even into the plane parallel to the surface, see Fig.~\ref{fig:ani}(c).

Supposing ferromagnetic order, the clusters under consideration clearly show uniaxial
magnetic anisotropy. 
The total anisotropy energy of the clusters as obtained from Eq.~\eqref{eq:anitot}, $\Delta E= \Omega(\vec{e}_x)-\Omega(\vec{e}_z)$, are listed in Table \ref{tab:explxcy}. For all clusters we find an easy-axis anisotropy, however, for the cluster L5C1 $\Delta E$ is quite small because of the in-plane contributions of the corner atoms as mentioned above. The MAE of the clusters L3C2 and L4C2 is more than four times larger than the MAE of the clusters
L3C1 and L4C1, breaking the rule of proportionality of $\Delta E$ to the number of magnetic atoms in the cluster ($N=7$ for C1 and $N=19$ for C2). This is, however, not surprising, since in case of the clusters  L3C1 and L4C1 the six corner atoms give a considerably decreased contribution to the MAE of the cluster as compared to the inner atom, while in case of the clusters  L3C2 and L4C2 this effect is reduced due to the larger number of inner atoms and also to the large contributions of the corner atoms, see Fig.~\ref{fig:ani}(a) and (b).


\begin{figure}[htb]
	\begin{center}
	\includegraphics[scale=1]{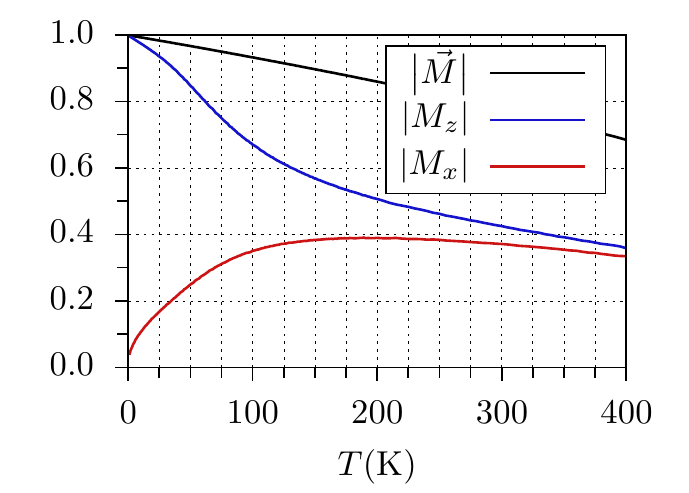}
	\end{center}
	\vspace{-0.5cm}
	\caption{Average magnetization and its components for cluster L4C2 as defined in Eqs.~\eqref{eq:tempmagt} and \eqref{eq:tempmaga}, respectively.}
	\label{fig:tempmagl4c2}
\end{figure}

Similar to the uncapped Co clusters, the capped clusters exhibit a nearly collinear ferromagnetic ground state. A slight non-collinearity is due to the DM interactions and the easy axes deviating from the $z$ directions. Because of the $C_{3v}$ symmetry of the clusters, the total magnetic moment points in the $z$ direction in the ground state. However, the ground state has a double degeneracy, related to the $z$ or $-z$ directions of the total moment.  

We evaluated the temperature dependent average magnetizations by MC simulations, where we used 
the parameters $T=400000$, $t_0=50$, $s=20000$, and for the Metropolis attempts we allowed any spin-direction over the unit sphere. 
The results are presented in Fig.~\ref{fig:tempmagl4c2} for the cluster L4C2. 
Due to the out-of-plane anisotropy $  | \vrd{M} |_{T=0} =   | M_z |_{T=0} \approx 1$, $  | M_x |_{T=0} = 0$, and in the high temperature limit all the directional averages are half of the total magnetization (see the uncovered case).

\begin{figure}[htb]
	\begin{center}
	\includegraphics[scale=1]{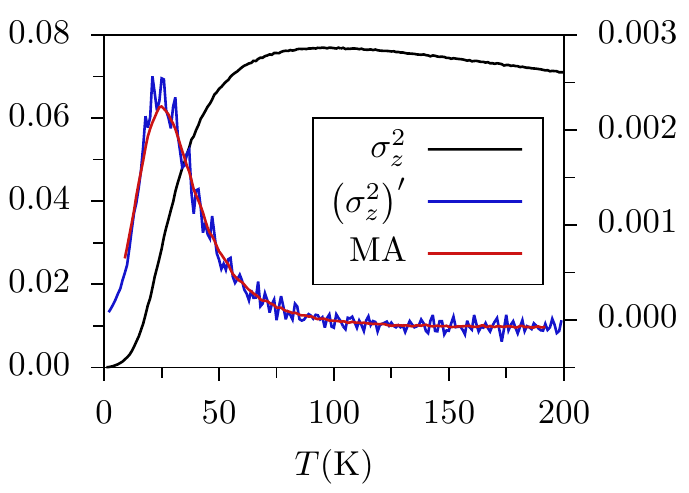}
	\end{center}
	\vspace{-0.5cm}
	\caption{MC variance $\sigma_z^2$ of the absolute value of the $z$ component of the magnetization in L3C1 with anisotropy set to $10.88\,\mathrm{meV}$, its temperature derivative $\left( \sigma_z^2 \right)^\prime$ and the moving average (MA) of the derivative. }
	\label{fig:ani4}
\end{figure}

In order to verify our concept of determining the blocking temperature from MC simulations as mentioned in context to Eq.~\eqref{eq:dev}, we performed a systematic study for cluster L3C1 by varying artificially the total MAE of the cluster.
The prescribed MAE was achieved by adding an appropriate amount of uniaxial on-site anisotropy uniformly at each site of the cluster.  In Fig.~\ref{fig:ani4} the results of such a model calculation are shown, where the total MAE of the cluster is set to 10.88\,meV. The parameters of the MC simulations were chosen $T=900000$, $t_0=50$ and $s=10000$, and no restriction was used for the trial spin-directions. As can be seen, the deviance $\sigma_z^2$ rapidly increases with increasing temperature, reaches a maximum plotted and then slightly decreases. The inflection point is determined by finding the maximum of its derivative, $\left( \sigma_z^2 \right)^\prime$. Since the derivative is very noisy, we evaluated the moving average (MA), where 15 temperature points were averaged. From the smooth MA curve it is easy to read out the temperature corresponding to the maximum point, $T_\sigma$. 

\begin{figure}[htb]
	\begin{center}
	\includegraphics[scale=1]{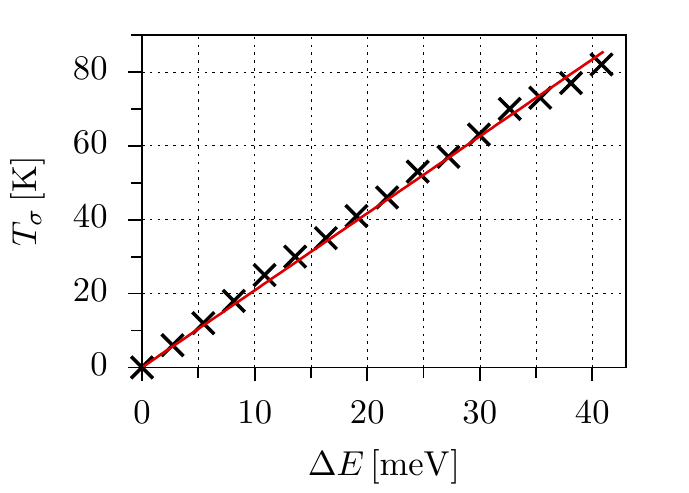}
	\end{center}
	\vspace{-0.5cm}
	\caption{Variance temperatures, $T_\sigma$, as a function of the MAE of the cluster L3C1.  The red line is a linear fit to the results with the slope $2.09\,\mathrm{K/meV}$.}
	\label{fig:aniall}
\end{figure}

We made further model calculations by setting the MAE of L3C1 to 16 different energies and specifying the inflection point (variance temperature) described above. We plotted $T_\sigma$ as the function of the total MAE of the system in Fig.~\ref{fig:aniall}. We found that $T_\sigma$ is proportional to the MAE, and the slope is 2.09 K/meV.  
Noticably, by increasing the simulation time $T_\sigma$ can be determined more accurately. 
For the considered clusters the corresponding results are summarized in Table \ref{tab:explxcy}. The ratio $T_\sigma/\Delta E$  is close to $2.09 \,\mathrm{K/meV}$ for most of the clusters. We note that the simulations lead to inaccurate results for cluster L5C1, because of the very small value of the MAE. 
Comparing with Eq.~\ref{eq:tb}, it is tempting to associate  $T_\sigma$ with the blocking temperature $T_B$ of superparamagnetic particles. 

In addition, we simulated the reversal mechanism by using a strategy similar as in Ref.~\onlinecite{Rozsa2014}. First the spins are set in random directions and then the system is thermalized. We accept the thermalization if $| M^z | > 0.6 \cdot | \vrd{M} |$, 
and count the steps after the thermalization, until the $z$ component of the magnetization does not reach $0.6 \cdot | \vrd{M} | $ in the opposite direction. The time of a single reversal is highly dependent on the initial conditions, so we measured it many times with different initial conditions. We determined the median value of the switching times, $\tau_{\rm med}$, instead of their average, because the latter one converges slower due to the Poisson distribution characteristic to the switching process. Moreover, $\tau_{\rm med}$ is proportional to the average value,  therefore, we associate $\tau_{\rm med}$ with $\tau_{\rm N}$, which just means the redefinition of $\tau_{0}$ in Eq.~\eqref{eq:neel}. According to our experience 1000 switchings are sufficient to achieve convergent value for $\tau_{\rm N}$, but in several cases we calculated 10000 reversals.
The time was measured in units of $N$ simple MC steps with $N$ being the number of spins. 

\begin{figure}[htb]
	\begin{center}
	\includegraphics[scale=1]{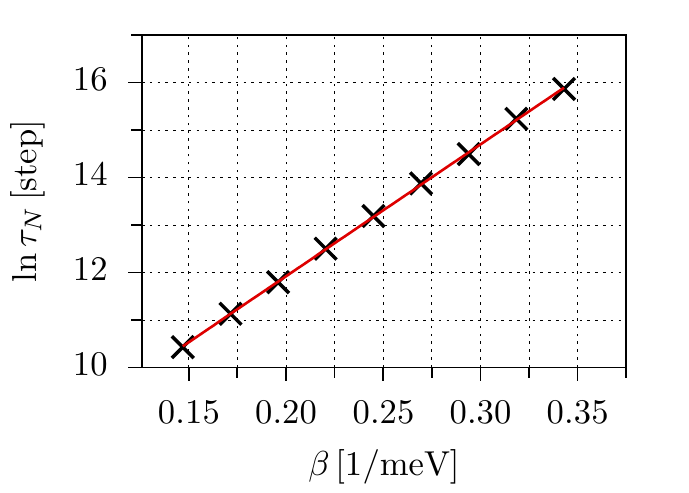}
	\end{center}
	\vspace{-0.5cm}
	\caption{Logarithm of the simulated switching time, $\tau_N$,  of cluster L4C2 as the function of the inverse temperature.  The red line shows a linear fit, with the slope $E_a=27.8\,\mathrm{meV}$.
	}
	\label{fig:neel}
\end{figure}

	\begin{table}[htb]
	\begin{center}
	\begin{tabular}{|c|c|c|c|c|}
	\hline
	Cluster&$\Delta E$ & $T_\sigma$ & $\displaystyle T_\sigma/\Delta E$ & $E_a$ \\  
	&  (meV)  &  (K)  &  (K/meV) &  (meV) \\ \hline
	L3C1&\ttb{5.9}&\ttb{13}&\ttb{2.19}&\ttb{6.3}\\
	L4C1&\ttb{6.3}&\ttb{15}&\ttb{2.37}&\ttb{6.9}\\
	L5C1&\ttb{0.67}&--&--&\ttb{1.95}\\
	L3C2&\ttb{25.7}&\ttb{53}&\ttb{2.06}&\ttb{26.7}\\
	L4C2&\ttb{26.4}&\ttb{54}&\ttb{2.05}&\ttb{27.8}\\
	L5C2&\ttb{25.9}&\ttb{53}&\ttb{2.05}&\ttb{27.2}\\
	\hline
	\end{tabular}
	\end{center}
	\caption{Magnetic anisotropy energy, $\Delta E$, according to Eq.~\eqref{eq:anitot}, variance temperature, $T_\sigma$, their ratio and the activation energy, $E_a$ obtained from the $\ln \tau_{\rm N}$ vs. inverse temperature curve, see Fig.~\ref{fig:neel}, for capped Co clusters. Note that for the cluster L5C1 the simulations were quite inaccurate because of the small MAE.}
	\label{tab:explxcy}
	\end{table}
	
We made the calculations for different  temperatures and, for the cluster L4C2, plotted the logarithm of the simulated N\'eel relaxation time in Fig.~\ref{fig:neel} as a function of the inverse temperature. It can clearly be seen that the data fit well to a straight line, therefore, the $\tau_{\rm N}$ indeed satisfies the N\'eel--Arrhenius law, Eq.~\eqref{eq:neel}. The slope of the logarithm equals the activation energy (energy barrier), in this case, $E_a=27.8 \, \mathrm{meV}$. We repeated the simulations of N\'eel times and determined the activation energies  for all the considered clusters covered by Au.  The results are summarized in the last column in Table \ref{tab:explxcy}. Apart from cluster L5C1, where we encountered difficulties in the simulations (see above), the activation energies are in good agreement with the total MAE of the clusters. From our simulations we, however, obtain that $E_a$ systematically overestimates  $\Delta E$, which indicates that the switching process doesn't perfectly correspond to a simple macrospin picture.

\section{Conclusions}
We applied the spin-cluster expension technique combined with the relativistic disordered moment picture \cite{Szunyogh2011} for finite-sized clusters and investigated how the parameters of an extended Heisenberg model vary by changing the size and the position of planar Co clusters on  Au(111) surface. The calculated parameters compare well with those for a Co monolayer, while some of the isotropic and DM interactions are larger between atoms at the perimeter. In case of Co clusters covered by Au we find large perpendicular magnetic anisotropy. Interestingly, however, for selected perimeter atoms the easy axis can turn to in-plane when the cluster is deposited on top of the surface.
 The presented method is capable to determine the parameters of more complex and magnetically frustrated systems, because there is no restriction to the geometry or to the magnetic ground state of the systems. 
 
We also studied the magnetism of the clusters at finite temperatures using  Monte Carlo simulations. 
We systematically investigated  the spin reversals of the covered clusters with perpendicular magnetic anisotropy. In terms of the variance of the magnetization in the easy direction we proposed a technique to determine the blocking temperature of superparamagnetic particles.
As expected,  the MAE of the clusters could be strongly correlated  with  the activation energy 
as deduced from  the N\'eel--Arrhenius law.

\section*{Acknowledgement}
Financial support for this work was provided by the National Research, Development and Innovation Office of Hungary under project No. K115575.


\bibliographystyle{apsrev4-1}
\bibliography{clusce_bib}

\end{document}